\documentclass[twocolumn,10pt,showpacs,preprintnumbers,amsmath,amssymb]{revtex4}
\usepackage{graphicx}
\usepackage{dcolumn}
\usepackage{bm}
\usepackage{color}
\usepackage{lineno,hyperref}
\usepackage{eqnarray,amsmath}
\def\bea {\begin{eqnarray}}
\def\eea {\end{eqnarray}}

\def\be {\begin{equation}}
\def\ee {\end{equation}}

\begin{document}
\title{Characterization of quark gluon plasma as seen through \\ the energy loss of heavy quarks}
\author{\it Ashik Ikbal Sheikh}

\author{ \it Zubayer Ahammed}
\email{za@vecc.gov.in}
\affiliation {Variable Energy Cyclotron Centre, HBNI, 1/AF Bidhan Nagar, Kolkata - 700 064, India }   
\begin{abstract}
The shear viscosity to entropy density ratio ($\eta/s$) of quark gluon plasma produced in ultra-relativistic heavy-ion collisions has been studied from the energy loss of heavy quarks in QGP  medium. We have also studied the bulk viscosity to entropy density ratio ($\zeta/s$) and fluidity measure ($F$) of the medium using the obtained $\eta/s$ values. In addition to that, we have estimated the heavy quark bound state potential ($V$) inside this medium. Our finding of $\eta/s$ agrees well with the results obtained by Lattice QCD (LQCD) and functional renormalization group technique.
\end{abstract}
\maketitle

\section{Introduction}
\label{intro}

  The ultra-relativistic heavy ion collisions in the Relativistic Heavy Ion Collider (RHIC) at BNL and Large Hadron 
Collider (LHC) at CERN aim to create a hot and densed deconfined state of QCD matter called Quark Gluon Plasma (QGP)\cite{BRAHMS,Muller12}. The main emphasis of these present day's heavy ion programes is to characterize the QGP or more precisely to determine it's transport coefficents\cite{Muller13}. Immediately after the creation, the QGP will be cooled by expansion due to the large internal pressure and will revert to the hadronic phase. The existence of these two phases (deconfined QGP phase and hadronic phase) has been confirmed by recent LQCD calculations\cite{Borsanyi12,Bazavov,Borsanyi14} and experimental observations\cite{br,ph,phn,str,al}. 
During the transition from the deconfined QGP phase to the hadronic phase, the system may encounter the critical point 
in the QCD phase diagram. The characterization of the medium at critical point is one of the most 
challenging problem in heavy ion collisions at the relativistic energies. The LQCD calculations indicate 
that the transition occurs at the critical temperature around $T_{c} \sim 155 $ MeV \cite{Karsch} at zero 
baryonic chemical potential.

   There are many probes in order to characterize the deconfined QGP phase. One of the efficient probe is the heavy quarks which are mostly 
produced from the fusion of partons in early stage of the collisions.  
No heavy quarks are produced at the latter stage if the temperature of the system is less than the mass of heavy quark pair and none in the hadronic matters. 
Hence, the total number of heavy quarks becomes frozen at the very beginning in the history of the 
collisions which  enables them to play a crucial role to characterize the QGP. 
Immediately after the production of heavy quarks, they will propagate through QGP medium and start losing energy via elastic collisions\cite{TG,BT,Alex,PP} and bremsstrahlung gluon radiations\cite{MG,mgm05,dokshit01,dead,wicks07,armesto2,B.Z,W.C,Vitev,AJMS}. There exists a transport parameter $\hat{q}$ which governs these collisional and radiative energy losses of the propagating heavy quarks\cite{Abhijit,Baier1,Baier2}. Generally, this $\hat{q}$ is sensitive to the interaction of the probes with the medium and has been used to calculate the shear viscosity to entropy density ratio($\eta/s$)\cite{Abhijit,Roy}. If a heavy quark anti-quark(Q$\bar{Q}$) bound state is present in this medium, then it's binding potential will be affected due to the temperature $T$ and $\eta/s$ of this medium\cite{Weber,Jorge,BKP1}. The effects of the medium are encoded in a temperature dependent $Q\bar{Q}$ bound state potential\cite{FK,BKP2,BKP3}. Hence, the estimation of the transport coefficients of QGP by using heavy quarks are of immense interest of research\cite{Abhijit,Baier1,Baier2,Roy,SKD2,Surasree}.

 In this article, we will revisit  the estimation of the various transport coefficients of the QGP. We have estimated $\eta/s$ from the $\hat{q}$ corresponding to the energy losses of heavy quarks in QGP. Along with that, we have also estimated bulk viscosity to entropy density ratio($\zeta/s$) and fluidity measure($F$) of QGP. It is interesting to note that, we found $\eta/s$ a satifactorily good agreement with the calculations of LQCD and functional renormalization group technique. In addition to that, Q$\bar{Q}$ bound state potential has been reported to understand the effect of the QGP medium we have characterized.
 
 This article is organised as follows: In the next section, we brifely discuss the energy loss of heavy quarks and transport 
parameter $\hat{q}$ associated to the energy loss. Here we consider the collisional energy loss of heavy 
quarks by Peigne and Pashier (PP) formalism\cite{PP} and the radiative energy loss by Abir, Jamil, Mustafa and Srivastava (AJMS)
 formalism\cite{AJMS}. In sec.\ref{sec3}, we discuss some properties of QGP, mainly the $\eta/s$, $\zeta/s$, $F$ and $Q\bar{Q}$ bound state potential. The summary and conclusion are in sec.\ref{sec4}.

\section{Energy loss and transport parameter}
\label{sec2}

In relativistic heavy ion collisions, the initially produced energetic heavy quarks will have to go through multiple 
scatterings and induced gluon bremsstrahlung radiations as it propagates through the medium. 
The calculation of the collisional energy loss per unit length $-dE/dx$ is reported by several authors\cite{TG,BT,Alex,PP}. 
One of the reliable calculation is made by Peigne and Pashier\cite{PP}. Another importnant and dominant procces of energy loss from a fast energetic parton in QGP is due to the gluon radiations. This radiative energy loss has also been calculated in the past(see Ref.\cite{MG,mgm05,dokshit01,dead,wicks07,armesto2,B.Z,W.C,Vitev,AJMS}). The energy loss by gluon emission has been calculated based on generalized dead cone approach in the Ref.\cite{AJMS}. We calculate these energy loss over the $QGP$ life time and finally avarage over the temperature evolution. The initial condition used for the hydrodynamic medium evolution as in Ref. \cite{my}. We take initial time $\tau_0 = 0.3$fm and freez-out time $\tau_f = 6$fm. In the context of these energy loss, the transport parameter $\hat{q}$ can be defined as the square of avarage momentum transfer between the probe and bath particles per unit length. If a highly energetic parton travels a distance $d$ in QGP so that it loses it's energy per unit length $-dE/dx$, then the $\hat{q}$ takes the form\cite{Baier2}:

 \begin{equation}
 \hat{q} =  \frac{1}{\alpha_sd} \left(- \frac{dE}{dx} \right),
\end{equation}

 where $\alpha_s$ is the strong coupling constant.

\begin{figure}[htb!]
\centering
\includegraphics[scale=0.33,angle = -90]{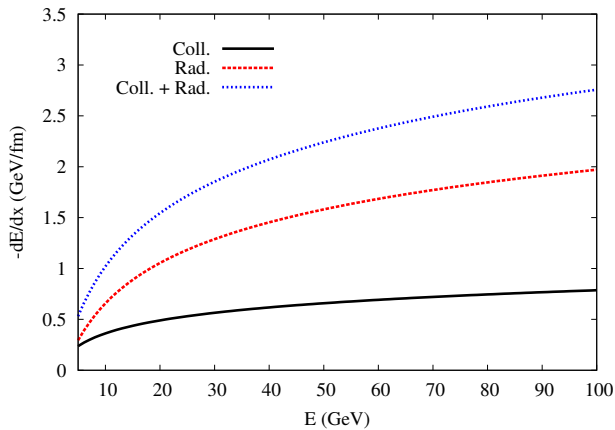}
\caption{The energy loss per unit length, $-dE/dx$, of a charm quark inside QGP medium as a function of its energy $E$, 
    obtained using PP formalism\cite{PP}(for Collisional loss (Coll.)) and AJMS formalism\cite{AJMS}(for Radiative loss (Rad.)).}
\label{dedx_c} 
\end{figure}

In Fig. \ref{dedx_c}, we show the various contributions of energy loss of a charm quark in QGP medium, obtained from \cite{PP,AJMS}. As the radiative procces is more dominating, the radiative energy loss is always larger than the collisional energy loss.

\begin{figure}[htb!]
\centering
\includegraphics[scale=0.33,angle = -90]{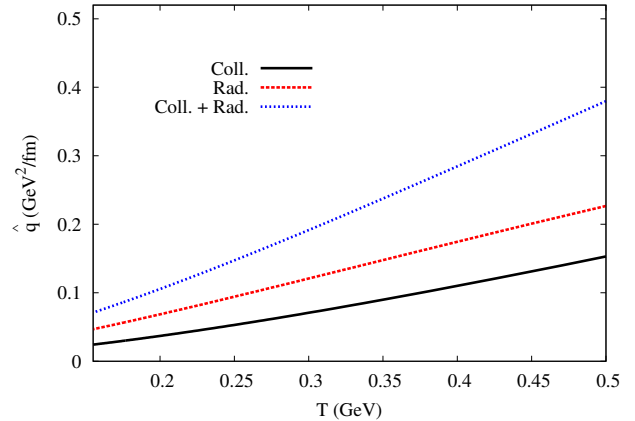}
\caption{Variation of transport parameter $\hat{q}$ with temperature $T$ of the QGP medium. We choose $d = 5$ fm.}
\label{TrPara}  
\end{figure}

 Fig. \ref{TrPara} displays variation of the transport parameter $\hat{q}$ as a function of temperature $T$ of QGP. The value of $\hat{q}$ increases with $T$. Since $\hat{q}$ is the measure of momentum transfer per unit length, 
larger the mementum transfer or energy loss, more is the value of $\hat{q}$.

\section{ Properties of QGP}
\label{sec3}
\subsection{Shear viscosity to entropy density \\ ratio ($\eta/s$)}

 Viscosity measures the resistance of a fluid deformed either by tensile stress or shear stress. The less viscosity means greater fludity. In order to characterize the QGP medium, amongst many,  $\eta/s$ is one of the important quantity. 
It is an important dimensionless measure of how imperfect or dissipative the QGP is. In this work, we evaluate $\eta/s$ by calculating the transport parameter $\hat{q}$ from the energy loss of  heavy quarks in QGP. 
A  heavy quark with certain momentum while propagating through QGP encounters the QGP bath particles and hence the momentum exchange occurs with the bath particles. 
We can define $\hat{q}$ by square of this avarage momentum exchange per unit length. 
So, the momentum of the energetic heavy quarks is shared by the low momentum (on an average) bath particles which 
causes minimization of momentum (or velocity) gradient in the system. Therefore, it s  related to the shear viscous coefficients of the system which drive the system towards a depleted velocity gradient. 

 With the definition of $\hat{q}$, the expression of $\eta/s$ reads as\cite{Abhijit}:
\begin{equation}
 \frac{\eta}{s} \approx  1.25 \frac{T^{3}}{\hat{q}} ,
\label{etaq}
\end{equation}
 where $T$ is the temperature of the QGP medium.

\begin{figure}[htb!]
\centering
\includegraphics[scale=0.42]{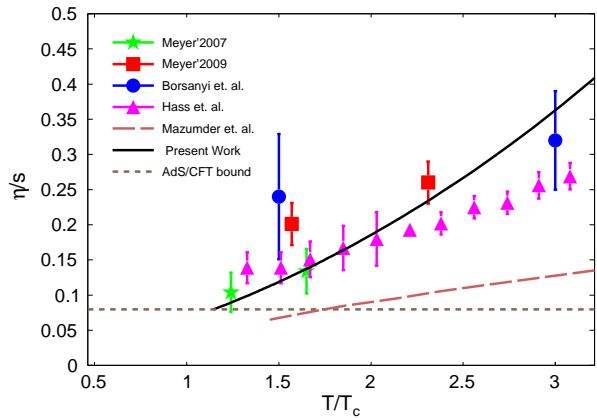}
\caption{The viscosity to entropy density ratio ($\eta/s$) as a function of $T/T_c$, is compared with the results obtained from charm quark's energy loss\cite{Surasree}, LQCD calculations\cite{Mayer07,Mayer09,BorsanyiLQCD14} and functional renormalization group calculations\cite{Haas}.}
\label{etaBys} 
\end{figure}

 Ads/CFT calculations\cite{AdsCFT} predics  a lower bound of $\eta/s$, which is $\frac{\eta}{s} \geq \frac{1}{4\pi}$. In Fig.\ref{etaBys} we display $\eta/s$ as a function of $T/T_c$ when the heavy quarks undergo both collisional and radiative processes. We compare $\eta/s$ with the results obtained by Mazumder et. al.\cite{Surasree} from charm quark's energy loss, LQCD calculations\cite{Mayer07,Mayer09,BorsanyiLQCD14} and functional renormalization group calculations\cite{Haas}. The value of $\eta/s$ by Mazumder et. al.\cite{Surasree} is lower than the LQCD calculations\cite{Mayer07,Mayer09,BorsanyiLQCD14} and calculations from functional renormalization group technique\cite{Haas}. Our finding of $\eta/s$ is in a good agreement with the LQCD calculations\cite{Mayer07,Mayer09,BorsanyiLQCD14} and calculations from functional renormalization group technique\cite{Haas}.

\subsection{Bulk viscosity to entropy density ratio ($\zeta/s$)}

 One another importnant quantity for QGP characterization is bulk viscosity to entropy density ratio ($\zeta/s$). Bulk viscosity acts as a resistance against the volume expansion of a fluid and it slows down the evolution of the system. A number of theoretical calculations have been done over the years to uncover the temperature dependence of $\zeta/s$ for the QGP.

\begin{figure}[htb!]
\centering
\includegraphics[scale=0.33,angle = -90]{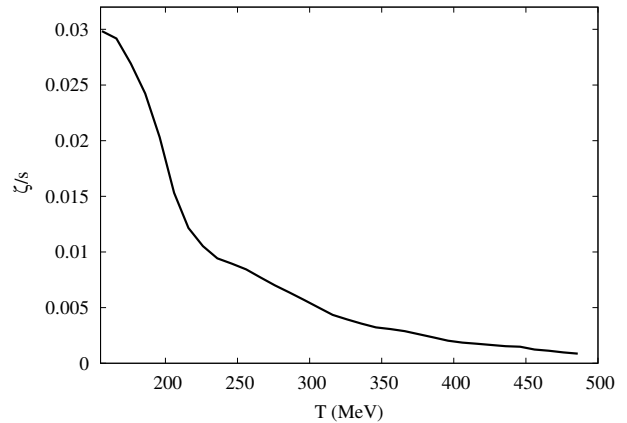}
\caption{Temperature ($T$) variation of bulk viscocity to entropy density ratio $\zeta/s$.}
\label{BulkVisco} 
\end{figure}
The bulk viscosity coefficient of QGP phase is parametrized as\cite{weinberge},

 \begin{equation}
 \frac{\zeta}{s} \approx 15\frac{\eta}{s}(\frac{1}{3} - c_{s}^{2})^2 ,
\end{equation}
 Here the values of $\eta/s$ are used from our findings(from Eq.\ref{etaq} ) and $c_s$, the velocity of sound is 
taken from LQCD calculations\cite{Borsanyi14}. In Fig.\ref{BulkVisco}, we show $T$ dependence of $\zeta/s$. $\zeta/s$ peaks near $T_c \sim 155$ MeV and decreases afterthat  as $T$ increases. The peak of $\zeta/s$ near $T_c$ is due to the lower value of $c_s$ around $T_c$ and large conformal breaking $(\frac{1}{3} - c_{s}^{2})^2$. $\zeta/s$  is insignificant at higher temperature and hardly plays any role in the high temteraure regime.

\subsection{Fluidity measure ($F$)}

 The measure of fluidity($F$) is based on the propagation of sound wave within the medium having dissipation due to visocity and under what condition the sound wave propagation has stopped beacuse of dissipation\cite{Liao}. Two length scales $L_{\eta}$ and $L_n$ in the medium have been introduced. $L_{\eta}$ provides a measure for the minimal wavelength of a sound wave to propagate in such a viscous medium and $L_n$ is the de-correlation length of a certain density-density spatial correlator which gives a natural scale of short range order in the system. In the most cases, this de-correlation length $L_n$ is the interparticle distance. Finally, the ratio of these two length scales gives a dimensionless quantity, termed as fluidity measure($F$). The expression of $F$ for a system having zero thermal conductivity and vanishing net charge density is taken from the Ref.\cite{Mafu} as:

\begin{figure}[htb!]
\centering
\includegraphics[scale=0.33,angle = -90]{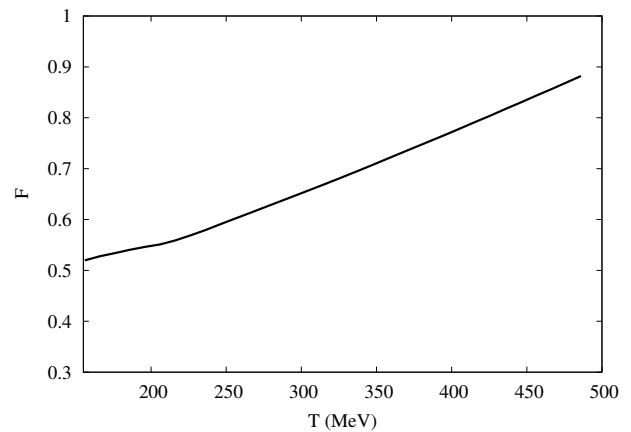}
\caption{Fluidity ($F$) as a function of temperature ($T$).}
\label{Fluidity} 
\end{figure}

\begin{eqnarray}
 F = \frac{\rho^{1/3}}{2\sqrt{2}}\Big[\zeta^2+8\zeta\eta+16\eta^2+2(P+\epsilon)\beta_0\zeta^2\left(\frac{\partial P}{\partial \epsilon}\right) \nonumber \\
+ 48\eta^2\beta_2(P+\epsilon)\left(\frac{\partial P}{\partial \epsilon}\right)\Big]^{1/2} \Big/ (P+\epsilon)\Big(\frac{\partial P}{\partial \epsilon}\Big)^{1/2}.
\end{eqnarray}

 Where $P$, $\rho$ and $\epsilon$ are the pressure, number density and energy density of the system respectively. $\Big(\frac{\partial P}{\partial \epsilon}\Big)^{1/2} = c_s$, is the velocity of sound. Using $P+\epsilon = sT$, $\beta_2 = \frac{3}{4P}$, $\frac{s}{P} = \frac{4}{T}$ and $\zeta = 0$ (with $s$ = entropy density of the system), we can arrive at

\begin{equation}
 F \equiv \frac{L_{\eta}}{L_n}.
\end{equation}
 Where,

 \begin{equation}
 L_{\eta} = \frac{1}{2\sqrt{2}}\frac{1}{T}\sqrt{16\frac{1}{c_{s}^2}\left(\frac{\eta}{s}\right)^2+144\left(\frac{\eta}{s}\right)^2}
\end{equation}
 and
\begin{equation}
 L_n = \left(\frac{4}{s}\right)^{1/3}.
\end{equation}

 Here $c_s$ and $s$ are used based on LQCD calculations\cite{Borsanyi14}.

 Fig. \ref{Fluidity} shows the fluidity measure($F$) of QGP as a function of $T$. Near $T_c$, the value of $F$ is less compare to other $T$ which implies that the QGP is more close to the perfect fluidity near $T_c$. As $T$ increases, $F$ increases which suggests that the system is deviating from near perfect fluidity.

\subsection{Heavy quark bound state $Q\bar{Q}$ potential ($V$)}

 In oreder to study the properties of the QGP medium, the $Q\bar{Q}$ bound state is one of the important probe. The $Q\bar{Q}$ bound states are more sensitive to the conditions taking place in the early stage of collisions and they have large masses and are not affected by the thermal medium. If this $Q\bar{Q}$ bound state has seperation $L$ at the vacuum, then the potential reads as\cite{Noronha}:

 \begin{equation}
 V = -\frac{1}{L}\frac{4\pi^2\sqrt\lambda}{\Gamma(1/4)^4} ,
\end{equation}
 where $\lambda$ is the 't Hooft coupling. Now this $V$ will be modified if the $Q\bar{Q}$ bound state is putted into a QGP medium of shear viscosity to entropy density ratio $\eta/s$ and temperature $T$. Then the medium modified potential looks like\cite{Noronha}:

 \begin{eqnarray}
 V = -\frac{2\sqrt\lambda}{L} \Big(\frac{\Gamma(3/4)}{\Gamma(1/4)}\Big)^2\Big[1-\frac{576\pi^2}{5} \frac{(LT)^4}{\eta\prime} \nonumber \\ 
 \times \frac{1}{(1+\eta\prime)^3} \Big(\frac{\Gamma(5/4)}{\Gamma(3/4)}\Big)^4 \Big] ,
\end{eqnarray}
 where $\eta\prime \equiv \sqrt{4\pi\eta/s}$ .

\begin{figure}[htb!]
\centering
\includegraphics[scale=0.33,angle = -90]{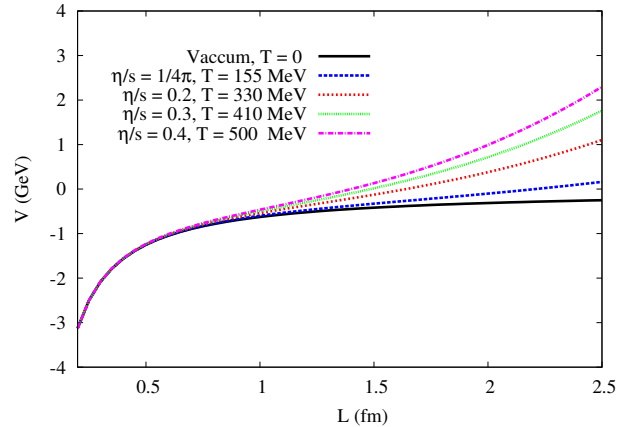}
\caption{Variation of heavy quark anti-quark bound state ($Q\bar{Q}$) potential $V$ with spatial seperation $L$  between $Q$ and $\bar{Q}$ in the bound state at different values of T and $\eta/s$.}
\label{PotL} 
\end{figure}

 In Fig.\ref{PotL}, we depict $Q\bar{Q}$ bound state potential $V$ as a function of spatial seperation $L$  between $Q$ and $\bar{Q}$ in the bound state for different values of $\eta/s$ and $T$ of the QGP medium. At vacuum($T=0$), $V$ is negative, i.e. the potential is attractive. As the medium comes into the picture, the nature of $V$ is modified. $V$ decreases (in magnitude) as the $\eta/s$ and $T$ increases, i.e. $V$ becomes weaker due to the presence of medium. More preciously, the medium screens the potential.


\section{Summary and Conclusion}
\label{sec4}
 
 In summary, we have estimated the transport parameter of heavy quarks propagating in the QGP medium. From the obtained transport parameter, we have estimated $\eta/s$, $\zeta/s$ and fluidity measure $F$ of the QGP medium. Our results  of $\eta/s$ are in good agreement with the LQCD and renormalization group calculations. In addition to that, the effect of this medium on the heavy quark bound state has also been discussed and found significant effects on it of the medium.

\section{Acknowledgement}
 
 A.I.S. acknowledges Mahfuzur Rahaman and Golam Sarwar for fruitful discussions.

\end{document}